\documentclass[preprint,showpacs, superscriptaddress, preprintnumbers,amsmath,amssymb]{revtex4}

\usepackage{booktabs}
\usepackage{color,graphicx}
\usepackage{amsmath}
\usepackage{dcolumn}
\usepackage{CJK}                 
\usepackage{bm}                  
\usepackage{soul}
\usepackage{enumerate}

\usepackage[dvipdfm,
            pdfstartview=FitH,
            CJKbookmarks=true,
            bookmarksnumbered=true,
            bookmarksopen=true,
            colorlinks,
            pdfborder=001,
            linkcolor=blue,
            anchorcolor=blue,
            citecolor=blue
            ]{hyperref}

\usepackage{float}

\begin{document}

\title{Effects of pairing, continuum and deformation on particles in the classically forbidden regions for Mg isotopes}

\author{Kaiyuan Zhang}
\affiliation{State Key Laboratory of Nuclear Physics and Technology,
School of Physics, Peking University, Beijing 100871, China}

\author{Dongyang Wang}
\affiliation{State Key Laboratory of Nuclear Physics and Technology,
School of Physics, Peking University, Beijing 100871, China}

\author{Shuangquan Zhang}
\email{sqzhang@pku.edu.cn}
\affiliation{State Key Laboratory of Nuclear Physics and Technology,
School of Physics, Peking University, Beijing 100871, China}

\begin{abstract}
  Particles in the classically forbidden regions are studied based on the deformed relativistic Hartree-Bogoliubov theory in continuum with PC-PK1 for magnesium isotopes. By analyzing the neutron and proton radii, it is found that the largest deviations from the empirical values appear at the predicted neutron halo nuclei $^{42}$Mg and $^{44}$Mg. Consistently, notable increases at $^{42}$Mg and $^{44}$Mg are found in the total number of neutrons in the classically forbidden regions that includes the number of neutrons in continuum. It is shown that the deformation effect, in general, increases the number of particles in the classically forbidden regions below the continuum threshold. The most deeply bound single-particle states play the dominant roles in the increase caused by deformation.
\end{abstract}

\date{\today}


\maketitle


\section{Introduction}

The development of radioactive ion beam facilities worldwide~\cite{ZHAN2010694c,GALES2010717c,THOENNESSEN2010688c} stimulates greatly the study of so-called exotic nuclei far from the $\beta$ stability line in both experimental side~\cite{Tanihata1995PPNP,Sorlin2008PPNP,Alkhazov2011583,TANIHATA2013215,SAVRAN2013210} and theoretical side~\cite{Vretenar2005PhysRep,Meng2006PPNP,Meng2015JPhysG,doi:10.1142/9872,Chatterjee201867}. Exciting discoveries in exotic nuclei including halo phenomena~\cite{Tanihata1985PRL}, pygmy resonances~\cite{PhysRevLett.95.132501}, and changes of magic numbers~\cite{PhysRevLett.84.5493,IWASAKI20007,KANUNGO200258,PhysRevLett.92.072502,PhysRevLett.96.012501,PhysRevLett.99.022503,PhysRevLett.100.152502,RN694,2019arXiv190707902T} have attracted a lot of attention.

As a microscopic quantum system, the atomic nucleus has many quantum characteristics and exhibits rich quantum phenomena. The quantum tunneling effect allows the penetration of nucleons in the classically forbidden (CF) regions and as a consequence has impacts on nuclear density distributions. For exotic nuclei that are very weakly bound systems, and particularly for halo nuclei that have very extended spatial density distributions, the tunneling effect is extremely important. Therefore, it is of particular interest to investigate particles in the CF regions for exotic nuclei.

In Refs.~\cite{Im2000PRC,Im2000}, particles in the CF regions for calcium isotopes were investigated with the spherical Skyrme Hartree-Fock theory. It was found that with increasing mass number $A$, the neutron number in the CF regions increases due to the increase of occupied neutrons in open shell while the proton number in the CF regions decreases due to proton orbits becoming more tightly bound. As the difference between the numbers of proton and neutron in the CF regions is very similar to the difference of the density distributions of proton and neutron, the particle number in the CF regions can give a signal for the halo or skin~\cite{Im2000}.

For exotic nuclei, pairing correlations and the coupling to continuum must be taken into account properly~\cite{DOBACZEWSKI1984103,Meng1996PRL}. Meanwhile, for open shell nuclei, one has to further deal with the deformation effect carefully. It is therefore necessary to investigate particles in the CF regions within a theoretical model that includes simultaneously the effects of pairing, continuum and deformation.

The covariant density functional theory (CDFT) describes nucleons as Dirac spinors which interact by exchanging effective mesons or point coupling in a microscopic and covariant way. The CDFT naturally includes the nucleonic spin degree of freedom and automatically results in a nuclear spin-orbit potential with empirical strength. It can give naturally the pseudospin symmetry in the nucleon spectrum~\cite{PhysRevLett.78.436,Meng1998Phys.Rev.C628,Meng1999Phys.Rev.C154,Chen2003Chin.Phys.Lett.358,Ginocchio2005PhysRep,Liang2015Phys.Rept.1} and the spin symmetry in the anti-nucleon spectrum~\cite{Zhou2003Phys.Rev.Lett.262501,Liang2015Phys.Rept.1}. Furthermore, the CDFT can include the nuclear magnetism~\cite{Koepf1989NPA}, which plays an important role in nuclear magnetic moments~\cite{Yao2006Phys.Rev.C24307,Arima2011,Li2011Sci.ChinaPhys.Mech.Astron.204,Li2011Prog.Theor.Phys.1185,Li2018Front.Phys.Beijing132109} and nuclear rotations~\cite{Meng2013Front,PhysRevLett.71.3079,Afanasjev2000NPA,PhysRevC.62.031302,PhysRevC.82.034329,Zhao2011Phys.Rev.Lett.122501,Zhao2011Phys.Lett.B181,Zhao2012Phys.Rev.C54310,Wang2017Phys.Rev.C54324,Wang2018Phys.Rev.64321}. Due to its successful description of many nuclear phenomena, the CDFT has become one of the most important microscopic methods in theoretical nuclear physics and has attracted wide attention~\cite{Ring1996,Vretenar2005PhysRep,Meng2006PPNP,Niksic2011PPNP,Meng2013Front,Meng2015JPhysG,doi:10.1142/9872}.

Based on the CDFT, taking advantage of the Bogoliubov transformation and solving the relativistic Hartree-Bogoliubov equations in the coordinate space, the relativistic continuum Hartree-Bogoliubov (RCHB) theory was constructed to consider pairing correlations and continuum in a unify and self-consistent way~\cite{Meng1996PRL,Meng1998NPA}. The RCHB theory has provided an interpretation of the halo in $^{11}$Li~\cite{Meng1996PRL}, predicted giant halos~\cite{Meng1998PRL,Meng2002Phys.Rev.C41302,Zhang2002Chin.Phys.Lett.312}, reproduced the interaction cross section and the charge-changing cross sections in light exotic nuclei in combination with the Glauber theory~\cite{Meng1998Phys.Lett.B1,Meng2002Phys.Lett.B209}, better restored the pseudo-spin symmetry in exotic nuclei~\cite{Meng1998Phys.Rev.C628,Meng1999Phys.Rev.C154}, and made predictions of exotic phenomena in hypernuclei~\cite{Lu2003Eur.Phys.J.A19} and new magic numbers in superheavy nuclei~\cite{Zhang2005Nucl.Phys.A106}.

In order to describe deformed nuclei properly, the deformed relativistic Hartree-Bogoliubov theory in continuum (DRHBc) was developed by solving the deformed relativistic Hartree-Bogoliubov equation in a Dirac Woods-Saxon basis~\cite{Zhou2010PRC,Li2012PRC,Li2012CPL}. The DRHBc theory was applied to study the chain of magnesium isotopes and an interesting shape decoupling between the core and the halo was predicted in $^{44}$Mg and $^{42}$Mg~\cite{Zhou2010PRC,Li2012PRC}. Later, the DRHBc theory was extended to incorporate the blocking effect which is required for the description of odd-nucleon systems~\cite{Li2012CPL}, and the density-dependent meson-nucleon couplings~\cite{Chen2012Phys.Rev.C67301}. Recently, using the DRHBc theory, the puzzles concerning the radius and configuration of valence neutrons in $^{22}$C were resolved and $^{22}$C was predicted to be a new candidate for deformed halo nuclei with shape decoupling effects~\cite{Sun2018PLB}.

In this work, the DRHBc theory is applied to investigate particles in the CF regions for magnesium isotopes. The results are compared with those from the RCHB theory and the RCHB calculations without pairing. The effects of pairing, continuum and deformation on particles in the CF regions are explored. Particles in the CF regions for a given single-particle state are also investigated.


\section{Theoretical framework} \label{theory}

The starting point of the CDFT is a Lagrangian density where nucleons are described as Dirac spinors which interact via exchange of effective mesons or point coupling. With the mean-field and the no-sea approximations, one can obtain the energy density functional for the nuclear system. According to the variational principle, one obtains the equation of motion for nucleons by minimizing the energy density functional with respect to the densities. Without pairing correlations, the relativistic Hartree equation can be solved either in the basis space or in the coordinate space~\cite{Ring1996,Zhou2003PRC,Ren2017Phys.Rev.C24313,ren2018stability}.

In the RCHB theory, the pairing is taken into account by the Bogoliubov transformation and the relativistic Hartree-Bogoliubov equations are solved in the coordinate space with spherical symmetry~\cite{Meng1996PRL,Meng1998NPA}. The details of the RCHB theory can be found in Ref.~\cite{Meng1998NPA}.

In the DRHBc theory, the potentials and densities are expanded in terms of the Legendre polynomials,
\begin{equation}\label{legendre}
f(\bm r)=\sum_\lambda f_\lambda(r)P_\lambda(\cos\theta),~~\lambda=0,2,4,\cdots,
\end{equation}
to include the deformation degree of freedom. The deformed relativistic Hartree-Bogoliubov equations have been solved in a Dirac Woods-Saxon basis, in which the radial wave functions have a proper asymptotic behavior in large $r$~\cite{Zhou2010PRC,Li2012PRC,Li2012CPL}. The details of the DRHBc theory can be found in Ref.~\cite{Li2012PRC}.

Following Ref.~\cite{Im2000}, the number of particles in the CF regions for a single-particle state $\beta$, $N_{\mathrm{CF}}^{\beta}$, is defined by
\begin{equation}\label{ncf}
  N_{\mathrm{CF}}^{\beta}=\int_{\bm R_{\mathrm{CF}}^{\beta}}^\infty \rho_{\beta}(\bm r) d^3\bm r,
\end{equation}
 in which $\bm R_{\mathrm{CF}}^{\beta}$ is the position where the mean field potential is equal to the single-particle energy $e_\beta$, i.e., $V(\bm R_{\mathrm{CF}}^{\beta})=e_{\beta}$, and $\rho_{\beta}$ is the density of the single-particle state $\beta$.

In the spherical RCHB theory, $N_{\mathrm{CF}}^{\beta}$ can be calculated by
\begin{equation}\label{spher}
\begin{split}
N_{\mathrm{CF}}^{\beta} & =\int_{\bm R_{\mathrm{CF}}^{\beta}}^\infty \rho_{\beta}(\bm r) d^3\bm r \\
& = \int_0^{2\pi} d\varphi \int_0^\pi \sin\theta d\theta \int_{R_{\mathrm{CF}}^{\beta}}^\infty r^2\rho_\beta(r) dr \\
& = 4\pi \int_{R_{\mathrm{CF}}^{\beta}}^\infty r^2\rho_\beta(r) dr,
\end{split}
\end{equation}
in which $R_{\mathrm{CF}}^{\beta}$ is the radial coordinate where the mean field potential is equal to the single-particle energy, i.e., $V(R_{\mathrm{CF}}^{\beta})=e_{\beta}$. The boundary of the CF regions is a spherical surface with radius $R_{\mathrm{CF}}^{\beta}$ correspondingly.

In the DRHBc model, the boundary of the CF regions is no longer a sphere but a surface determined numerically. For every fixed $\theta$, there exists a radial position $R_{\mathrm{CF}}^{\beta}(\theta)$, and they satisfy
\begin{equation}\label{boundary}
V[R_{\mathrm{CF}}^{\beta}(\theta),\theta]=e_\beta.
\end{equation}
As a result, $N_{\mathrm{CF}}^{\beta}$ is calculated by
\begin{equation}\label{defor}
\begin{split}
N_{\mathrm{CF}}^{\beta} & =\int_{\bm R_{\mathrm{CF}}^{\beta}}^\infty \rho_{\beta}(\bm r) d^3\bm r \\
& = \sum_\lambda \int_0^{2\pi} d\varphi \int_0^\pi P_\lambda(\cos\theta)\sin\theta \left[\int_{R_{\mathrm{CF}}^{\beta}(\theta)}^\infty r^2\rho_\lambda^\beta(r) dr \right]d\theta \\
& = 2\pi\sum_\lambda\int_0^\pi P_\lambda(\cos\theta)\sin\theta \left[\int_{R_{\mathrm{CF}}^{\beta}(\theta)}^\infty r^2\rho_\lambda^\beta(r) dr \right]d\theta .
\end{split}
\end{equation}

It is noted that the states in continuum have energies higher than the threshold, which means they cannot be occupied from the classical perspective, therefore the number of particles in continuum, $N_{\mathrm{continuum}}$, should contribute also to the number of particles in the CF regions. Finally, the number of particles in the CF regions is calculated by
\begin{equation}\label{Ncf}
N_{\mathrm{CF}} =\sum_\beta v_\beta^2 \cdot d_\beta \cdot N_{\mathrm{CF}}^{\beta},
\end{equation}
where $v_\beta^2, d_\beta$ are the occupation probability and the degeneracy of single-particle state $\beta$, respectively. The summation runs over all of the occupied single-particle states. For the bound states, $N_{\mathrm{CF}}^{\beta}$ is calculated by Eq.~(\ref{spher}) or Eq.~(\ref{defor}), and for the states in continuum, $N_{\mathrm{CF}}^{\beta}=1$.

In the following, we use $N_{\mathrm{CF}}^n$ ($N_{\mathrm{CF}}^p$) to represent the neutron (proton) number in the CF regions that includes the neutron (proton) number in continuum, and $N_{\mathrm{CF}}^n-N_{\mathrm{continuum}}^n$ to represent the number of neutrons in the CF regions below the continuum threshold. The sum of $N_{\mathrm{CF}}^n$ and $N_{\mathrm{CF}}^p$ gives the number of total nucleons in the CF regions for a nucleus.


\section{Numerical details}\label{numerical}

To explore the effects of pairing, continuum and deformation, the even-even Mg nuclei from proton drip line to neutron drip line are investigated using the DRHBc theory and the RCHB theory. For the particle-hole channel, the relativistic density functional PC-PK1~\cite{Zhao2010Phys.Rev.C54319}, which has turned out to be very successful in providing good descriptions of the isospin dependence of the binding energy along either the isotopic or the isotonic chain~\cite{Zhao2012Phys.Rev.C64324,Zhang2014Front.Phys.529}, is adopted. For the particle-particle channel, the pairing is taken into account by using a density-dependent zero-range pairing force~\cite{Meng1998NPA,Li2012PRC}. In the present DRHBc calculations, we fix the box size $R_{\mathrm{box}}=20$ fm, the mesh size $\Delta r=0.1$ fm, and the angular momentum cutoff $J_{\mathrm{max}}=23/2~\hbar$. The parameters in pairing force follow Ref.~\cite{Li2012PRC}, i.e., the saturation density $\rho_{\mathrm{sat}}=0.152~\mathrm{fm}^{-3}$, and the pairing strength $V_0=-380~\mathrm{MeV}~\mathrm{fm}^3$ together with the cutoff energy of $60$~MeV for the pairing window. It was shown that these numerical conditions give results accurately enough in energies and radii for light nuclei~\cite{Zhou2003PRC,Li2012PRC}. The maximal order $\lambda_{\mathrm{max}}=4$ is adopted in the Legendre expansion Eq.~(\ref{legendre}) of the deformed potentials and densities. For determination of the Woods-Saxon basis an energy cutoff $E^+_{\mathrm{cut}}=300$ MeV for positive-energy states is used,  and the number of negative-energy states in the Dirac sea is the same as that of positive-energy states in the Fermi Sea~\cite{Zhou2003PRC}. The numerical details in the RCHB calculations are the same as those in Ref.~\cite{Xia2018ADNDT}.


\section{Results and Discussion}\label{results}

\begin{figure}[htbp]
  \centering
  \includegraphics[scale=0.6,angle=0]{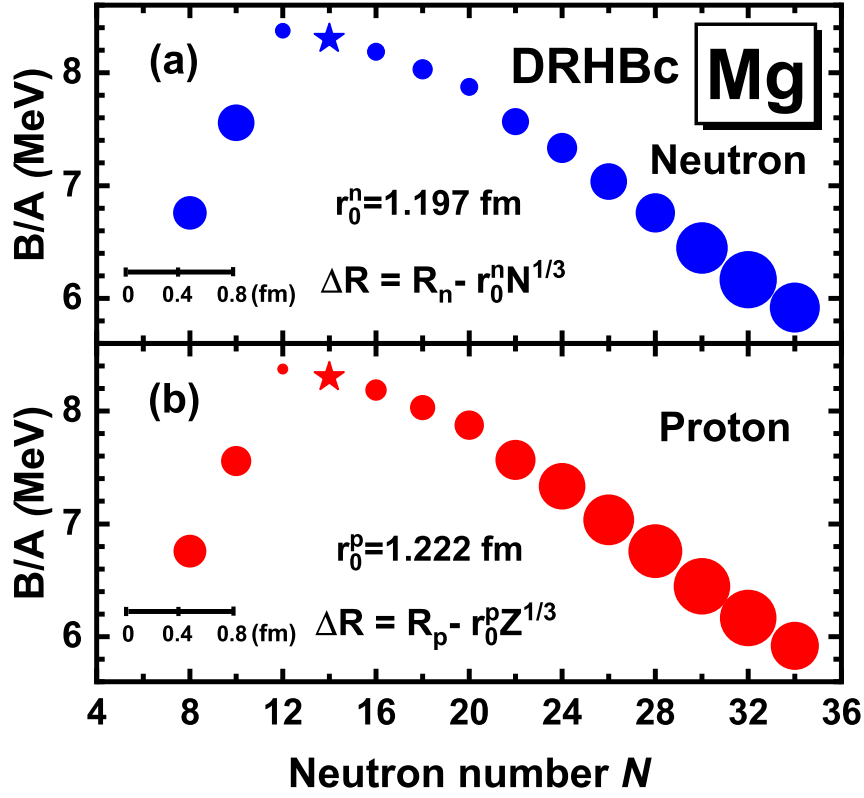}
  \caption{Binding energies per nucleon from the DRHBc calculations for Mg as functions of the neutron number. (a) The size of each point is proportional to the deviation between the calculated neutron root-mean-square radius $R_n$ and the empirical value $r_0^n N^{1/3}$, in which $r_0^n=1.197$ fm, determined from $^{26}$Mg (star), is the lowest ratio $R_n/N^{1/3}$ in the Mg isotopic chain. (b) Same as (a), but for protons, with the empirical value $r_0^p Z^{1/3}$, in which $r_0^p=1.222$ fm, also determined from $^{26}$Mg (star).}
\label{fig1}
\end{figure}

In Fig.~\ref{fig1}, the binding energies per nucleon for magnesium isotopes from the DRHBc calculations are shown as functions of the the neutron number. It can be seen that the binding energies per nucleon decrease as the neutron number moves away from $12$, i.e. $^{24}$Mg, which is the most stable nucleus in Mg isotopes. The deviations between the calculated neutron (proton) root-mean-square (rms) radii $R_n (R_p)$ and the empirical values $r_0^n N^{1/3} (r_0^p Z^{1/3})$ denoted by the size of each point are shown. $r_0^n$ and $r_0^p$ are the lowest ratios $R_n/N^{1/3}$ and $R_p/Z^{1/3}$ determined from $^{26}$Mg (star). In general, as the nucleus moves away from $^{26}$Mg, the deviation from the empirical value becomes larger for both neutron and proton. For neutron, the largest deviations from the empirical values appear at $^{42}$Mg and $^{44}$Mg; this is consistent with the suggested halo phenomena in Refs.~\cite{Zhou2010PRC,Li2012PRC}. For protons, the empirical value $r_0^p Z^{1/3}$ is a constant, and after $^{26}$Mg the proton radius increases gradually except for $^{46}$Mg. Instead of the increase in neutron radius from  $^{44}$Mg to $^{46}$Mg, the proton radius slightly decreases, as $^{44}$Mg has a prolate shape whereas $^{46}$Mg is spherical in the DRHBc calculations.

\begin{figure}[htbp]
  \centering
  \includegraphics[scale=0.4,angle=0]{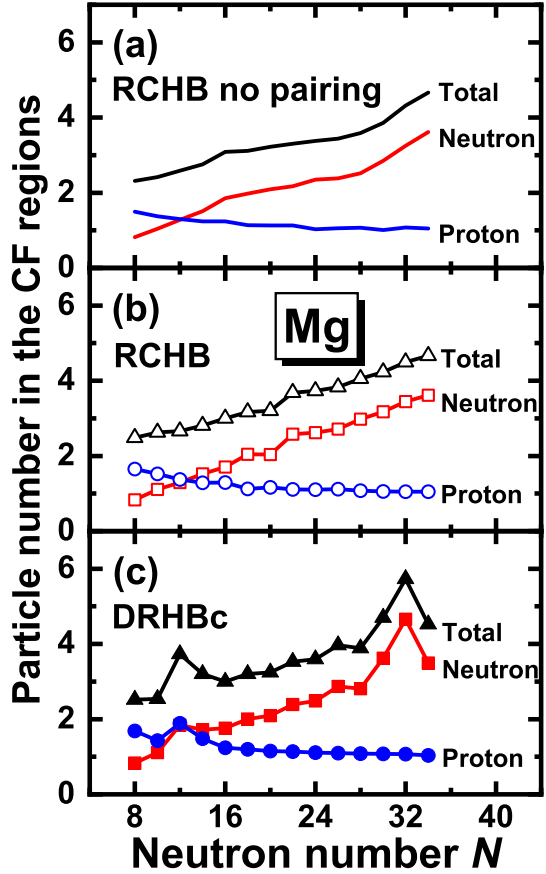}
  \caption{The number of neutrons, protons and total nucleons in the CF regions from (a) the RCHB calculations without pairing, (b) the RCHB calculations, and (c) the DRHBc calculations as functions of the neutron number for magnesium isotopes.}
\label{fig2}
\end{figure}

It was suggested that particles in the CF regions can also provide information on the appearance of halos or skins~\cite{Im2000,Im2000PRC}. Figure~\ref{fig2} shows the number of neutrons, protons and total nucleons in the CF regions as functions of the neutron number for magnesium isotopes. In general, with the neutron number, the evolution trends from different calculations are consistent. The number of protons in the CF regions ($N_{\mathrm{CF}}^p$) gradually decreases with the neutron number, whereas the number of neutrons in the CF regions ($N_{\mathrm{CF}}^n$) rapidly increases. As a result, the number of total nucleons in the CF regions also increases with the neutron number. Similar results have been found in a previous study with the Skyrme Hartree-Fock model~\cite{Im2000}. For the nuclei near the neutron drip line, the difference $N_{\mathrm{CF}}^n-N_{\mathrm{CF}}^p$ becomes as large as $2$. $N_{\mathrm{CF}}^n$ increases with the neutron number due to the increase in occupied neutrons in weakly bound single-particle levels whereas $N_{\mathrm{CF}}^p$ decreases due to proton orbits becoming more tightly bound, which is the reason for the large difference and has been analyzed carefully in Ref.~\cite{Im2000}. Although consistent in general, there are some differences in the results from the DRHBc theory vs the RCHB theory. Notable increases can be seen in $N_{\mathrm{CF}}^n$ from the DRHBc calculations at the predicted neutron halo nuclei $^{42}$Mg~\cite{Li2012PRC} and $^{44}$Mg~\cite{Zhou2010PRC}. Moreover, there is a prominent bump near $^{24}$Mg for both $N_{\mathrm{CF}}^p$ and $N_{\mathrm{CF}}^n$.

\begin{figure}[htbp]
  \centering
  \includegraphics[scale=0.6,angle=0]{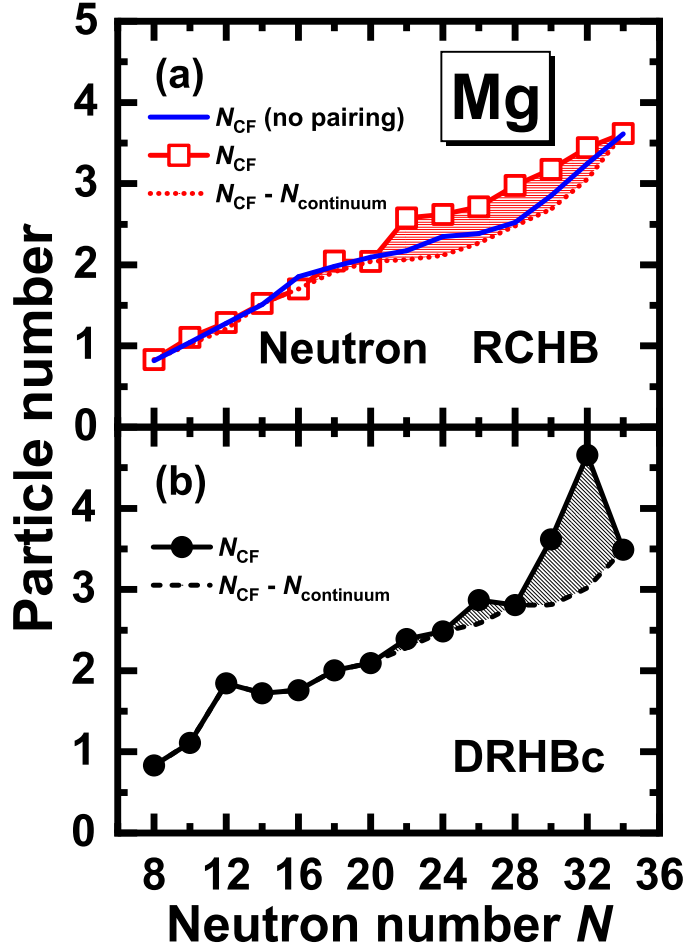}
  \caption{The neutron numbers in the CF regions from (a) the RCHB calculations and the RCHB calculations without pairing, and (b) the DRHBc calculations as functions of the neutron number for magnesium isotopes. The neutron numbers in the CF regions below the continuum threshold from the RCHB calculations (dotted line) and the DRHBc calculations (dashed line) are also shown for comparison. Shaded regions represent neutrons in continuum.}
\label{fig3}
\end{figure}

As for halo nuclei, occupation of the continuum becomes very important~\cite{DOBACZEWSKI1984103,Meng1998NPA}, to study the continuum effect, Fig.~\ref{fig3} shows the number of neutrons in the CF regions, the number of neutrons in the CF regions below the continuum threshold ($N_{\mathrm{CF}}^n-N_{\mathrm{continuum}}^n$), and the contributions from the continuum obtained from the RCHB and DRHBc calculations. As seen in Fig.~\ref{fig3}(a) and (b), for nuclei with $8\le N\le 20$, the numbers of neutrons in continuum are very small or even $0$, but for the neutron-rich side, the numbers of neutrons in continuum are remarkable. In panel (a), $N_{\mathrm{CF}}^n$ from the RCHB calculations without pairing are also shown for comparison. For isotopes with $8\le N\le 20$, $N_{\mathrm{CF}}^n$ values from the RCHB calculations without pairing are very close to $N_{\mathrm{CF}}^n$ and $N_{\mathrm{CF}}^n-N_{\mathrm{continuum}}^n$ from the RCHB calculations, but for neutron-rich nuclei, $N_{\mathrm{CF}}^n$ from the RCHB calculations without pairing are in between them. This is because for stable isotopes with $8\le N\le 20$, neutrons occupy the well bound single-particle levels which are slightly affected by pairing correlations, but on the neutron-rich side, pairing correlations can scatter valance neutrons into continuum, which significantly contributes to $N_{\mathrm{CF}}^n$. In panel (b), the notable increases in $N_{\mathrm{CF}}^n$ at $^{42}$Mg and $^{44}$Mg in the DRHBc calculations can be well understood now. It shows that the increases mainly come from the increases in neutron numbers in continuum. This phenomenon is consistent with the previous predictions of halo phenomena in the nuclei $^{42}$Mg~\cite{Li2012PRC} and $^{44}$Mg~\cite{Zhou2010PRC}.

\begin{figure}[htbp]
  \centering
  \includegraphics[scale=0.4,angle=0]{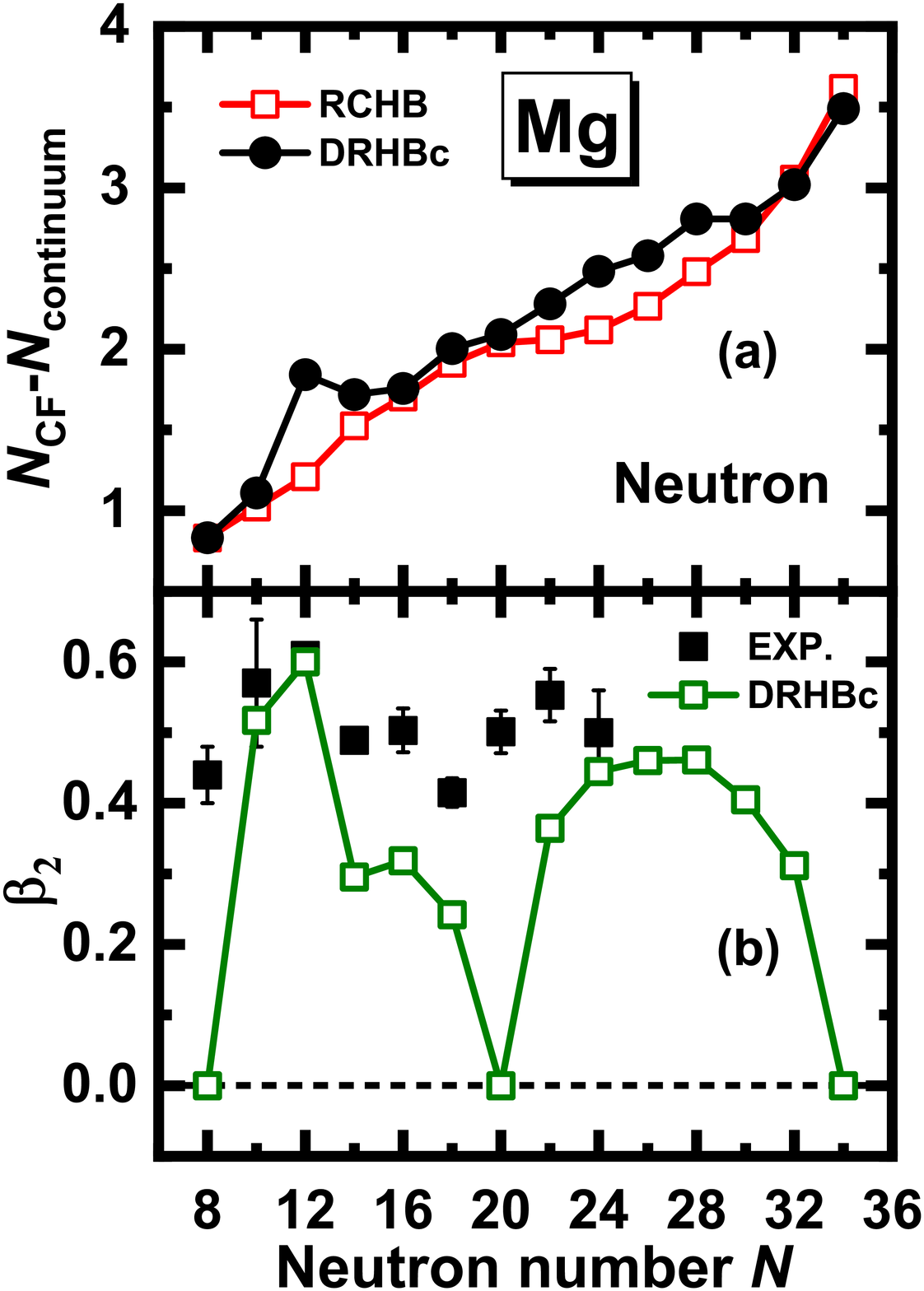}
  \caption{(a) The number of neutrons in the CF regions below the continuum threshold calculated by the RCHB theory and the DRHBc theory as functions of neutron number for magnesium isotopes. (b) The ground state quadrupole deformation $\beta_2$ for magnesium isotopes as a function of the neutron number within the DRHBc calculations. Experimental values (``EXP.") are taken from Ref.~\cite{2016ADNDT}.}
\label{fig4}
\end{figure}

In order to see the possible deformation effect, for Mg isotopes, the number of neutrons in the CF regions below the continuum threshold, $N_{\mathrm{CF}}^n-N_{\mathrm{continuum}}^n$, and the ground state deformations from the DRHBc calculations are shown in Fig.~\ref{fig4}. As seen in Fig.~\ref{fig4}(b), the ground states of $^{20}$Mg, $^{32}$Mg and $^{46}$Mg are spherical in the DRHBc calculations. As the neutron number increases from $8$ to $20$ and from $20$ to $34$, the shape of the nucleus goes from spherical to prolate, and then back to spherical. The ground state of $^{24}$Mg is well deformed with the largest quadrupole deformation $\beta_2\approx 0.60$, which is consistent with the experimental value $\beta_2=0.609(6)$~\cite{2016ADNDT}. Comparing (a) and (b), one finds that the appearance of the bump near $^{24}$Mg is caused by the strong deformation effect. Moreover, Mg isotopes with $20<N<34$ are also deformed with large deformations $\beta_2\approx0.3\sim0.5$, and $N_{\mathrm{CF}}^n-N_{\mathrm{continuum}}^n$ values from the DRHBc theory are larger than those from the RCHB theory in this region except for $^{44}$Mg. For $^{44}$Mg, as can be seen in Fig.~\ref{fig3}(b), the coupling to continuum is extremely strong and more than $1.5$ neutrons are scattered into continuum in the DRHBc theory (only $\sim0.5$ neutron is scattered into continuum in the RCHB theory).

To better understand the deformation effect on the number of particles in the CF regions, taking $^{24}$Mg as an example, constraint DRHBc calculations with $\beta_2=0.1, 0.2, 0.3, 0.4, 0.5$ and $0.6$ are performed. Here pairing correlations are not taken into account to exclude the influence of the pairing effect. It is found that both the neutron number and the proton number in the CF regions increase with increasing $\beta_2$ and the increases in $N_{\mathrm{CF}}^n$ and $N_{\mathrm{CF}}^p$ are very close. From $\beta_2=0.1$ to $\beta_2=0.6$, the numbers of neutrons in the CF regions are $1.331, 1.393, 1.542, 1.639, 1.761$ and $1.847$, respectively. To be more explicit, the contributions of different single-neutron states are further investigated. Each state is labeled with $\Omega_i^\pi$ where $\Omega$ is the projection of the angular momentum on the symmetry axis, $\pi$ is the parity, and $i$ is used to order the level in the $\Omega^\pi$ block, as $\Omega$ and $\pi$ are good quantum numbers in the axially deformed system. It is found that the increase in $N_{\mathrm{CF}}^n$ mainly (more than $85\%$) comes from the contributions of the two most deeply bound single-neutron states $(1/2)_1^+$ and $(1/2)_1^-$, whereas for other occupied states, the neutron numbers in the CF regions increase or decrease very little.

\begin{figure}[htbp]
  \centering
  \includegraphics[scale=0.4,angle=0]{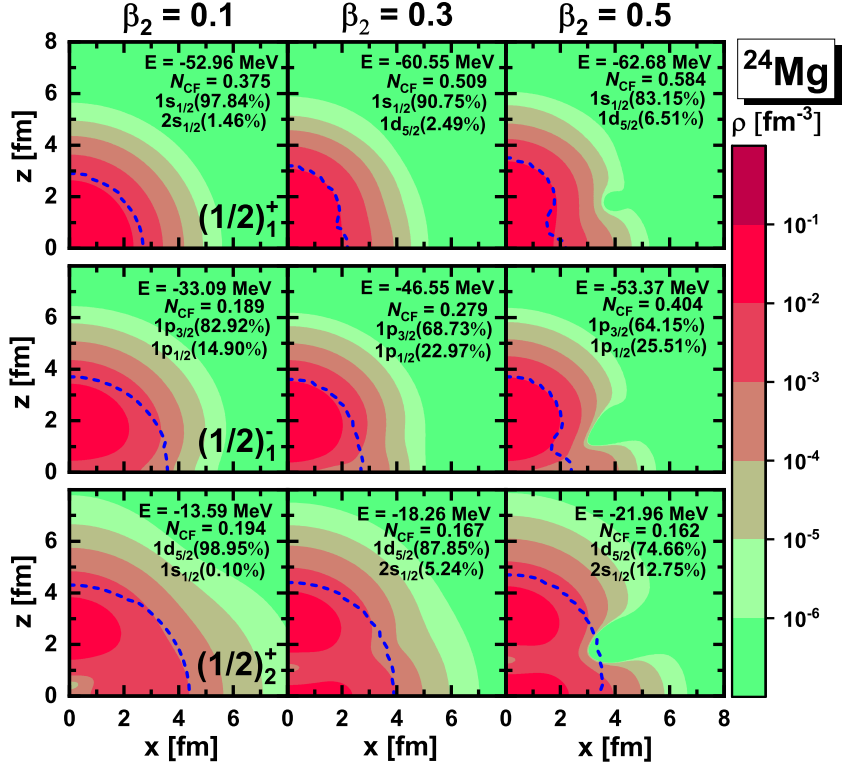}
  \caption{For $^{24}$Mg, density distributions and the boundary of the classically forbidden regions (dashed line) of single-neutron states $(1/2)_1^+$, $(1/2)_1^-$ and $(1/2)_2^+$ from the constraint DRHBc calculations without pairing at the quadrupole deformation parameters $\beta_2=0.1, 0.3$ and $0.5$. The $z$ axis is the symmetry axis. In each panel, the corresponding single-neutron energy, the number of neutrons in the classically forbidden regions and the two largest spherical components are shown.}
\label{fig5}
\end{figure}

Figure~\ref{fig5} shows the density distributions and the boundary of the classically forbidden regions for the single-neutron states $(1/2)_1^+$, $(1/2)_1^-$ and $(1/2)_2^+$ obtained from the constraint DRHBc calculations at $\beta_2=0.1, 0.3$ and $0.5$. In each panel, the corresponding single-neutron energy, the neutron number in the classically forbidden regions and the two largest spherical components are also given. It can be seen that the three states become more bound with increasing deformation. For the states $(1/2)_1^+$ and $(1/2)_1^-$, the numbers of neutrons in CF regions increase by more than $0.2$, but for the state $(1/2)_2^+$, the number of neutrons in CF regions decreases only by about $0.03$. When the deformation increases, as seen in Fig.~\ref{fig5}, on one hand, the intersection point between the boundary of the CF regions and the $z$ axis moves away from the center, whereas that between the boundary and the $x$ axis approaches the center, and as a result, the CF regions expand with the deformation. On the other hand, the density distributions of neutron states also evolve with the deformation as a result of the changes in the mean-field potential, the single-neutron energies and the component mixing. For the most deeply bound states $(1/2)_1^+$ and $(1/2)_1^-$, not only do the CF regions expand, but also the neutron densities at the boundary of the CF regions increase a bit, and then the numbers of neutrons in the CF regions increase a lot and dominate the increase in $N_{\mathrm{CF}}^n$. For other states like $(1/2)_2^+$, the boundaries of CF regions are farther away from the center and the density distributions are more diffuse, therefore the deformation effect on them is weak and their numbers of neutrons in the CF regions slightly change with deformation. Then it is concluded that, in general, the deformation will increase the particle number in the CF regions below the continuum threshold and the most deeply bound single-particle states play the dominant roles in this effect.

\begin{figure}[htbp]
  \centering
  \includegraphics[scale=0.37,angle=0]{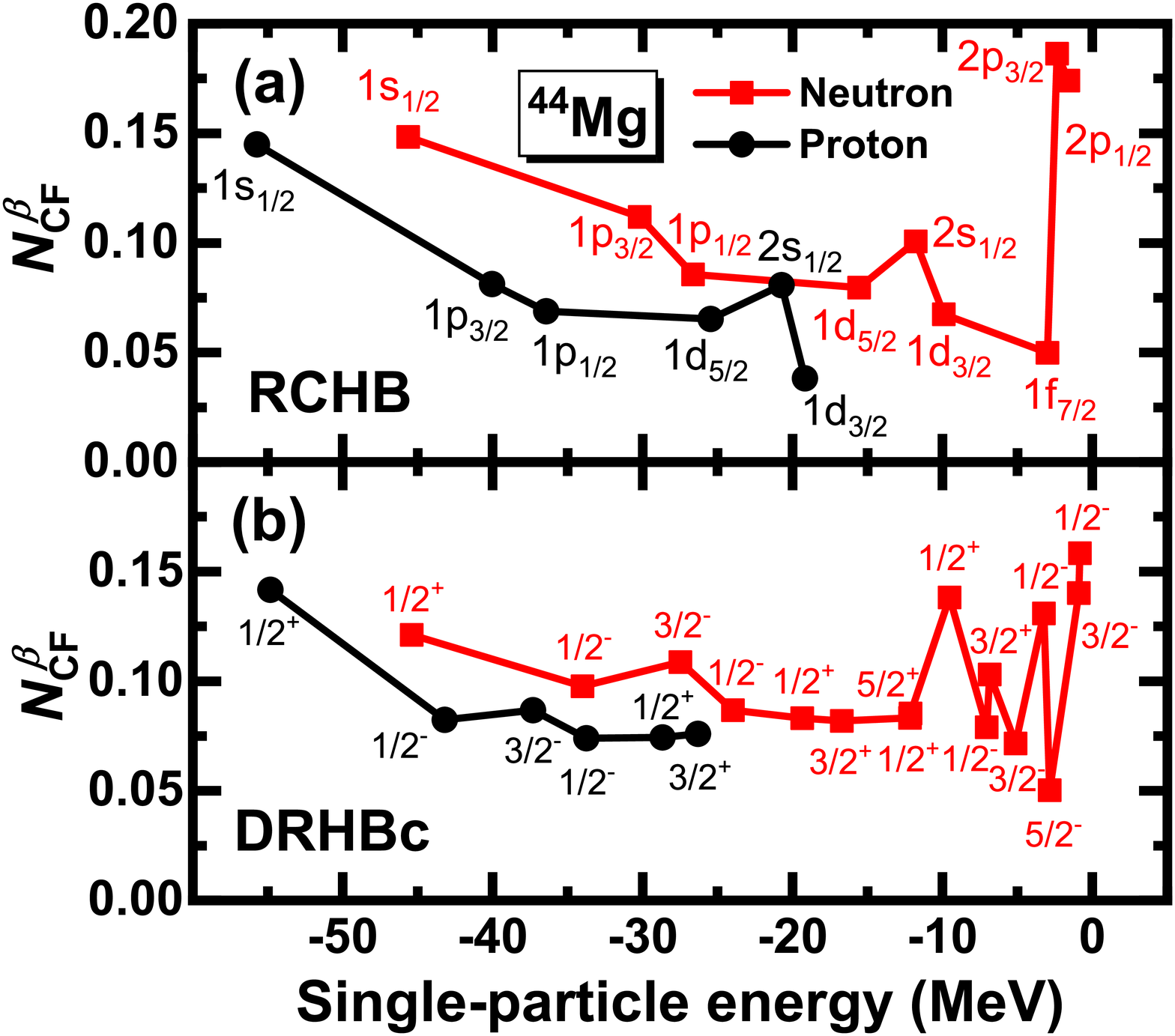}
  \caption{The number of particles in the CF regions for each bound single-particle state, $N_{\mathrm{CF}}^\beta$, from (a) the RCHB calculations and (b) the DRHBc calculations as functions of the single-particle energy for $^{44}$Mg $(N=32)$.
  }
\label{fig6}
\end{figure}

To further study the relation between the halo phenomenon and particles in the CF regions, taking the neutron-rich nucleus $^{44}$Mg as an example, the numbers of particles in the CF regions for each bound single-particle state, $N_{\mathrm{CF}}^\beta$, from the RCHB calculations and the DRHBc calculations as functions of the single-particle energy are shown in Fig.~\ref{fig6}. A similar trend of $N_{\mathrm{CF}}^\beta$ can be found for neutrons and protons, as they share the same mean field self-consistently. However, one cannot find a monotonic relationship between $N_{\mathrm{CF}}^\beta$ and the single-particle energy $e_\beta$ in either theoretical calculations. In Fig.~\ref{fig6}(a), by comparing the $N_{\mathrm{CF}}^\beta$ with the corresponding orbital angular momentum $l$ of the single-particle state, one can find a smaller $N_{\mathrm{CF}}^\beta$ for the single-particle state with a larger $l$. This feature was also seen in Ref.~\cite{Im2000} as a state with large $l$ has a high centrifugal barrier. A weakly bound single-particle state with a low orbital angular momentum $l$ is helpful in the formation of halos~\cite{Meng1996PRL,Meng1998PRL}. It is shown that weakly bound $2p$ states have a larger $N_{\mathrm{CF}}^\beta$ than the others.

In Fig.~\ref{fig6}(b), the evolutions of $N_{\mathrm{CF}}^\beta$ in the DRHBc calculations are similar to those in the RCHB calculations. It can be seen, the $(1/2)_5^-$ state and the $(3/2)_3^-$ state, that are near the threshold, have the largest $N_{\mathrm{CF}}^\beta$. These two weakly bound orbitals were suggested to give rise to the formation of the halo in $^{44}$Mg~\cite{Zhou2010PRC} and we find the $p$-wave components for the states $(1/2)_5^-$ and $(3/2)_3^-$ are $69.6\%$ and $81.5\%$, respectively. On the other hand, the weakly bound states $(5/2)_1^-$ and $(3/2)_2^-$ have much smaller $N_{\mathrm{CF}}^\beta$, as their dominant components are $f$ waves ($97.4\%$ and $80.1\%$, respectively) that corresponds to a high centrifugal barrier. Therefore, it is concluded that a weakly bound single-particle state with a low centrifugal barrier has a larger $N_{\mathrm{CF}}^\beta$.


\section{Summary}\label{summary}

In summary, particles in the classically forbidden regions for magnesium isotopes are investigated within the relativistic continuum Hartree-Bogoliubov theory and the deformed relativistic Hartree-Bogoliubov theory in continuum with PC-PK1. It is found that, with an increasing neutron number, the number of neutrons in the classically forbidden regions rapidly increases, whereas the number of protons in the classically forbidden regions gradually decreases. Particles in the classically forbidden regions for a given single-particle state have been investigated and it is shown that a weakly bound single-particle state with a low centrifugal barrier has more particles in the classically forbidden regions. Similar conclusions have been obtained by the spherical Skyrme Hartree-Fock theory in Ref.~\cite{Im2000}.

In this paper, the effects of pairing, continuum and deformation on particles in the classically forbidden regions are studied for the first time. For the stable isotopes, as neutrons occupy the well bound single-particle levels, the neutron numbers in the classically forbidden regions are only slightly affected by pairing correlations, but on the neutron-rich side, pairing correlations can scatter valance neutrons into continuum, which significantly contributes to neutron numbers in the classically forbidden regions. Including neutrons in continuum, notable increases in the total number of neutrons in the classically forbidden regions at $^{42}$Mg and $^{44}$Mg have been found; this is consistent with the largest deviations from empirical values of radii at $^{42}$Mg and $^{44}$Mg shown in Fig.~\ref{fig1} and the predictions of halo phenomena in $^{42}$Mg~\cite{Li2012PRC} and $^{44}$Mg~\cite{Zhou2010PRC}. Comparing the results from the DRHBc theory and the RCHB theory, it is found that, in general, the deformation effect enhances the number of neutrons in the classically forbidden regions below the continuum threshold. The most deeply bound single-particle states play the dominant roles in the increase caused by deformation.

\begin{acknowledgments}

The authors are indebted to J. Meng for constructive guidance and valuable suggestions. Helpful discussions with L. S. Geng, Z. X. Ren, S. B. Wang, Y. Zhang, B. Zhao and P. W. Zhao are greatly acknowledged. This work was partly supported by the National Science Foundation of China (NSFC) under Grants No.~11875075, No.~11621131001, No.~11935003 and No.~11975031, and the National Key R\&D Program of China (Contract No.~2018YFA0404400, No.~2017YFE0116700).

\end{acknowledgments}


\end{document}